\documentclass[prl,amsmath,amssymb, twocolumn, showpacs]{revtex4}

\usepackage[dvips]{graphicx}

\usepackage{epsfig}

\usepackage{amsmath}

\begin{document}

\title{Hastening, delaying, or averting sudden death of quantum entanglement}

\author{A.\ R.\  P. Rau$^{1*}$, Mazhar Ali$^2$ and G.\ Alber$^2$}
\affiliation{$^1$Department of Physics and Astronomy, Louisiana State University,
Baton Rouge, Louisiana 70803, USA \\
$^2$ Institut f\"{u}r Angewandte Physik, Technische Universit\"{a}t Darmstadt, D-64289, Germany}

\begin{abstract}The finite time end of entanglement between two decohering qubits can be modified by local, unitary actions performed during the decoherence process. Depending on the time when such action is taken, the end can be speeded up or slowed down, or even averted all together. This phenomenon offers practical applications for the stabilization of entangled quantum states. Details concerning hastening or 
delaying the finite time end of entanglement are presented for two qubits which decay spontaneously into statistically independent reservoirs.
\end{abstract}


\maketitle

Entanglement is a key feature of the quantum physics of more than one particle. From its historical beginnings \cite{EPR} to the current practical interest in it as a core resource for the field of quantum information sciences \cite{Nielsen}, this property of quantum systems continues to fascinate and to shed new light onto the nature of our quantum world. The fields of quantum computing, quantum cryptography and key distribution \cite{QKD}, and quantum teleportation\cite{teleport} all rely on having entangled states of two qubits. Since each qubit is inevitably subject to decoherence and decay processes, no matter how much they may be screened from the external environment, it is important to consider possible degradation of any initially established entanglement. In particular, there has been increasing discussion of what has been called ``sudden death", a finite time when the entanglement disappears even under decoherence mechanisms which may be only asymptotic in time \cite{end, Eberly, Almeida}. Clearly, such finite time disappearance of entanglement can seriously affect its application in any of the above fields.

It is well known in the context of spontaneous emission processes \cite{spontaneous} or delayed choice experiments \cite{delayed},
for example, that characteristic quantum phenomena and effects of decoherence can be influenced significantly
by suitable actions, such as measurements. Therefore,
it would be of interest if A(lice) and B(ob), the two members of the entangled pair, can take suitable individual actions when faced with the prospect of loss of entanglement to postpone that end. 
Some studies on changing the initial state into an equivalently entangled but more robust state have been carried
out \cite{initialization}. 
In this Letter, we deal with the more direct question that, even given an initial state and a set up which will end in disentanglement at finite time, can they themselves take suitable actions later to change the fate of their entanglement.
We answer in the affirmative. In particular, it is shown that simple
local unitary operations
can alter the time of disentanglement. This is even possible if these
local operations are separated spacelikely so that they are not connected by any causal relation.
The operations we consider can either hasten or delay that time depending on its time of application.
A suitable window for this application can even avert completely the finite or sudden death. In that case, entanglement will persist and decay only asymptotically just as do the decoherences for both qubits. While our discussion will be for two qubits, it is clear that similar results will apply also to other systems such as qubit-qutrit \cite{Ann} and qutrit-qutrit \cite{Lastra, Derkacz} where also questions of the finite end of entanglement have been considered. We will present such results elsewhere.

We consider the model of two two-level atoms and
``amplitude damping" in the form of spontaneous (pure exponential) decay into statistically
independent reservoirs from the excited to the ground state being the only postulated dynamics \cite{Eberly}. Whatever the initial state, whether pure or mixed, and whether entangled (non-separable) or not, the final state reached in asymptotic time is clearly one of both atoms in the ground state, that is, a product state of the two ground-state atoms/qubits with no entanglement. It is also a pure state with zero entropy. The much discussed model \cite{Eberly} considers mixed states with a density matrix of the form

\begin{equation}
\rho(t) = \frac{1}{3} \left( 
\begin{array}{cccc}
a(t) & 0 & 0 & 0 \\ 
0 & b(t) & z(t) & 0 \\ 
0 & z(t) & c(t) & 0 \\
0 & 0 & 0 & d(t)
\end{array}
\right).      
\label{eqn1}
\end{equation}
The coefficients $(a, b, c, d, z)$ may be considered real, and Tr $\rho=(a+b+c+d)/3 =1$. The four states are described as usual by $(|++\rangle, |+-\rangle, |-+\rangle, |--\rangle)$ with Alice and Bob's states shown as the first and second entries, respectively, in the ket, and $+/-$ denoting excited/ground state. The initial condition chosen, of $b(0)=c(0)=z(0)=1$, along with the only evolution, that + decays to $-$ at a steady rate $\exp(-\Gamma t/2)$ in amplitude, keeps $b(t)=c(t)$ throughout.

The form of $\rho(t)$ in Eq.~(\ref{eqn1}) is preserved by time evolution. 
The off-diagonal density matrix element is given by $\dot{z}(t)=-\Gamma z(t)$
(an overhead dot indicates differentiation with respect to time) 
and the diagonal ones by
\begin{equation}
\frac {d}{dt}
\left( 
\begin{array}{c}
\rho_{++} \\ 
\rho_{+-} \\ 
\rho_{-+} \\ 
\rho_{--}
\end{array}
\right)  =  \left( 
\begin{array}{cccc}
-2\Gamma & 0 & 0 & 0 \\ 
\Gamma & -\Gamma & 0 & 0 \\ 
\Gamma & 0 & -\Gamma & 0 \\
0 & \Gamma & \Gamma & 0
\end{array}
\right)  \left( 
\begin{array}{c}
\rho_{++} \\ 
\rho_{+-} \\ 
\rho_{-+} \\
\rho_{--}
\end{array}
\right).
\label{eqn2}
\end{equation}
No elaborate derivation is necessary, these equations having an obvious structure dictated by the ``decay" from + to ``feed" into $-$. Their solutions are also immediate. In terms of a logarithmic, dimensionless time parameter, $\gamma = \exp(-\Gamma t/2)$, we have
 
\begin{eqnarray}
\rho_{++}(t) =a(t) & = & a(0)\gamma^4 \nonumber \\
\rho_{+-}(t) =\rho_{-+} (t) =b(t) & = & [b(0)+a(0)]\gamma^2-a(0)\gamma^4 \nonumber \\
\rho_{--}(t) =d(t) \!& = &\!\! 3+a(0)(\gamma^4 \!- \!\gamma^2)\!-\!\![3\!-\!d(0)]\gamma^2 \nonumber \\
z(t) & = & z(0)\gamma^2. 
\label{eqn3}
\end{eqnarray}

Were $(b, c, z)$ to be the only non-zero elements in Eq.~(\ref{eqn1}), we would have an entangled pure state $(|+-\rangle+|-+\rangle)$. The coefficients would all decay with a factor $\gamma^2$, and so would the entanglement only asymptotically. The additional choice of either $(a(0)=1, d(0)=0)$ or $(a(0)=0, d(0)=1)$ gives a mixed state which is non-separable. Both choices give the same entropy, defined as $-\sum \rho_i \ln \rho_i$ in terms of the eigenvalues of $\rho$. This value is $\ln (3/4^{1/3})$ at $t=0$ and decreases to 0 asymptotically when the system is in the pure state $|--\rangle$. But their evolution of entanglement or separability is very different \cite{Eberly, Zubairy}. 

Various measures of entanglement all coincide in their conclusion that the second choice of $d(0)=1$ leads to non-separability only asymptotically at infinite time whereas the first choice with $a(0)=1$ leads to a finite time for the end of entanglement, the ``sudden death" \cite{Eberly}. Concurrence \cite{Wootters} having been discussed in previous papers, we choose {\it negativity} as a more easily measurable quantity in terms of the partially transposed density matrix \cite{Horodecki} as an indicator of non-separability.

The {\it negativity} is defined as the sum of the absolute values of all the negative eigenvalues of the partially transposed density matrix \cite{vidal} for a quantum state. For the $2 \otimes 2$ system, there can be at most one such possible negative eigenvalue \cite{sanpera}. Viewing Eq.~(\ref{eqn1}) in terms of $2 \times 2$ blocks and transposing the off-diagonal blocks to define such a partial transpose, its eigenvalues contain one that can possibly be negative. This eigenvalue, or alternatively, six times that value, is given by $a(t)+d(t)-\sqrt{[a(t)-d(t)]^2+4z^2(t)}$, and it can take negative values so long as $ad<z^2$. When $a(0)=0$, since $a(t)$ from Eq.~(\ref{eqn3}) remains zero at all times, the system retains non-separability for all finite t. 
However, for the choice $(d(0)=0, a(0)=1)$, the system starts as non-separable or entangled but when $a(t)d(t)$ crosses $z^2(t)$ during the subsequent evolution, the entanglement is lost. It can be seen that the time $t_0$ at which this happens is 

\begin{equation}
\Gamma t_0 = \ln (1/\gamma_0^2)=\ln (1+1/\sqrt{2}).
\label{eqn4}
\end{equation}
This is the time of ``sudden death" \cite{Eberly}. At this point, we have $a=6-4\sqrt{2}, b=2\sqrt{2}-2, d=1, z=2-\sqrt{2} $.

Previous papers have examined the evolution of entanglement for different initial choices of the above parameters \cite{Eberly, Zubairy}. The evolution of Werner states \cite{Werner}, with a form slightly different from the one in Eq.~(\ref{eqn1}) with non-zero entries in the other two corners as well, has also been studied \cite{Zubairy}. It has also been noted that different ``initializations", wherein an initial given state such as in Eq.~(\ref{eqn1}) is switched to another with equivalent entanglement, can lead to a change in the time of non-separability \cite{initialization}. A recent paper has collected compactly necessary and sufficient conditions for this under both amplitude and phase damping \cite{Huang}. 

Another observation, with multimode radiation fields, is that spontaneous emission can also lead to a revival of entanglement from a separable configuration \cite{multimode}. In three-level atoms or qutrits, finite end of entanglement for pairs and abrupt changes in lower bounds on entanglement have been noted \cite{Lastra}, as also quantum interference between different decay channels creating an asymptotic entanglement \cite{Derkacz}. The finite end phenomenon has also been noted for mixed qubit-qutrit states \cite{Ann} and it seems to be a generic phenomenon for all entangled-pair systems. 

We turn, however, to a different question, whether, given the qubit-qubit system above and initial conditions that lead to separability in finite time, a suitable intervention may alter that time. Such a question is clearly even of practical interest because, as noted in \cite{Roszak}, ``finite end may affect the feasibility of solid-state based quantum computing". Therefore, a simple intervention that prolongs the entanglement resource can be of broad interest. Indeed, the above discussion and, especially, the asymmetry noted between the two choices of whether it is $a$ or $d$ that is initially zero, suggests a way for such an intervention. When $a(t)=0$, the non-separability in the mixed state because of the presence of $d$ simply continues as the states that are entangled ``decay down" to enhance $d$. The other situation of $a(0)=1$ and, therefore, a non-zero $a(t)$ is quite different, because it feeds into the entangled sector ``from above". At a crucial point/time when $ad>z^2$, which, as is especially clear from the partial transposed density matrix (see Eq.~(\ref{eqn1})), has to do with the $(|++\rangle, |--\rangle)$ sector, it is this feed which swamps the entangled sector and makes the mixed state separable.

\begin{figure}
\scalebox{2.0}{\includegraphics[width=1.7in]{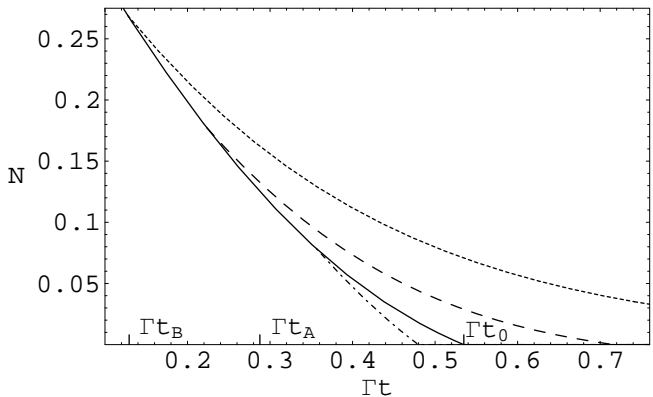}}
\caption{Evolution of negativity $N$ with time for an initial mixed state in Eq.~(\ref{eqn1})
with $a(0)=b(0)=c(0)=z(0)=1, d(0)=0$. The solid line shows the undisturbed evolution, with sudden death at
$\Gamma t_0 \approx 0.5348$. Other dotted and dashed lines show the effect of switching the values of $a$ and $d$ at different times. Those after $\Gamma t_A \approx 0.2877$ hasten, and those before delay, the sudden death. Switches earlier than $\Gamma t_B \approx 0.1293$ avoid sudden death altogether, negativity vanishing only asymptotically in time.}
\label{fig1}
\end{figure}

\begin{figure}
\scalebox{2.0}{\includegraphics[width=1.7in]{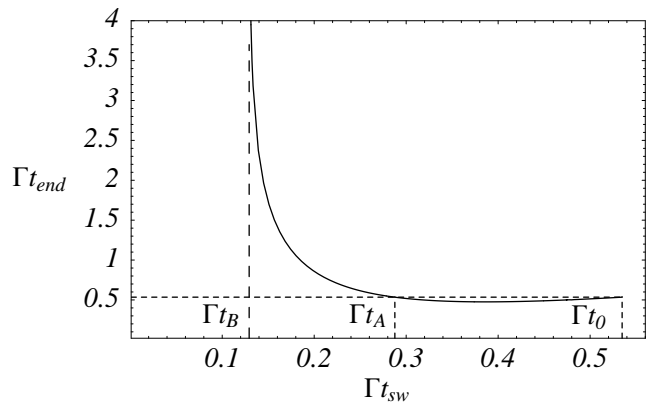}}
\caption{The time for the end of entanglement is plotted against the time of switching $+$ and $-$ in Eq.~(\ref{eqn1}).
Starting on the right at switching times of $\Gamma t_0 \approx 0.5348$,
the curve has a broad and small dip before rising rapidly to infinite time at $\Gamma t_B \approx 0.1293$.}
\label{fig2}
\end{figure}

The above diagnosis of which aspect of the asymmetry between $a$ and $d$ is responsible for 
the separability suggests a ``switch" between them. Such a switch, which leaves the other coefficients $(b,c,z)$ unchanged, amounts to interchanging + and $-$ for both qubits. This is a local unitary transformation that both Alice and Bob can easily implement, by individual $\sigma_x$ operations for spins or laser coupling of the excited and ground states for two-level atoms. Consider the same initial condition as before, with $(a(0)=1, d(0)=0)$, which leads to sudden death. Before the time $t_0$
corresponding to the end of entanglement,
consider such local unitary operations that merely interchange $a$ and $d$.
If this is done at the time $t_A$ when $a=d$, which happens when
${\rm exp}(-\Gamma t_A) \equiv \gamma_A^2 =3/4$,
clearly there will be no effect upon the subsequent
evolution, the end still coming at time $t_0$. See Figs. \ref{fig1} and \ref{fig2}.
If the switch is made at any time intermediate between $t_A$ and $t_0$ (see FIG. \ref{fig2}), separability occurs earlier, a minimum being at the switch time $\Gamma t_1 \approx 0.357$. 

More interestingly, a switch earlier than $t_A$ prolongs the entanglement as shown in the figures. 
Moreover, switch times before $t_B$ with $\Gamma t_B \approx 0.1293$ avoid the finite time end all together,
leading to separability only asymptotically. As a practical matter, therefore, when Alice and Bob separate
at $t=0$ and know that they face an end to their entanglement at $t_0$, they can agree beforehand to make the local unitary switch between + and $-$ at a certain time as desired to alter that end.

In FIG \ref{fig1}, {\it negativity} is plotted against the parameter $\Gamma t$.
The solid line corresponds to a situation when no switch is made and the sudden death happens at $\Gamma t_0 \approx 0.5348$.
If the switch is made at $\Gamma t_1 \approx 0.357$ (dotted-dashed line),
the sudden death reaches a minimum value of $\Gamma t_< \approx 0.48$.
Any switch made earlier than $t_A$ leads to a delay in sudden death in comparison with $t_0$.
Two instances are shown in the upper curves.
If the switch is made at $\Gamma t \approx 0.223$ (dashed line),
the {\it negativity} comes to an end at $\Gamma t \approx 0.716$. 
Any switch made earlier than $\Gamma t_B \approx 0.1293$ avoids a finite end, leading only to asymptotic decay of entanglement.

FIG \ref{fig2} displays the time of sudden death $t_{end}$ against the time of switching $t_{sw}$.
The earlier the switch is made than $\Gamma t_A \approx 0.2877$, the more the end of entanglement is delayed.
Sudden death is avoided completely when the switch takes place earlier than $\Gamma t_B \approx 0.1293$. 

\begin{figure}
\scalebox{2.0}{\includegraphics[width=1.7in]{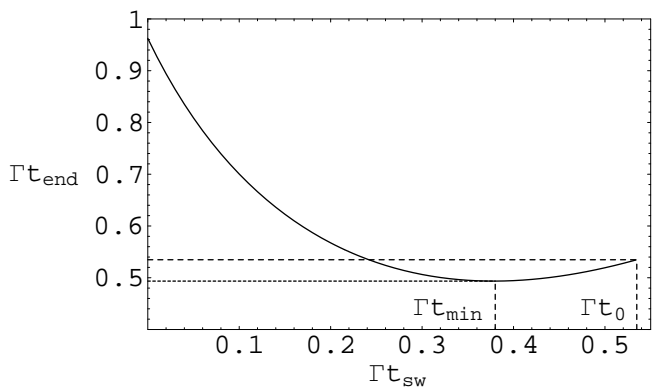}}
\caption{As in FIG. \ref{fig2} when the switch is done at only one end. Note that the maximum delay, which occurs at $t_{sw}=0$, is now finite.}
\label{fig3}
\end{figure}

Interestingly, switching + and $-$ at only one end, that is, either Alice or Bob makes the local unitary $\sigma_x$ transformation, also alters the end of entanglement. Now, $a$ and $c$ in the density matrix in Eq.~(\ref{eqn1}) are interchanged, as also $b$ and $d$, while $z$ moves to the corners of the anti-diagonal. The roles of $(a,d)$ and $(b,c)$ in Eq.~(\ref{eqn3}) are interchanged and we find that sudden death is hastened or delayed depending on the time of switching but it is now no longer averted indefinitely. FIG. \ref{fig3} shows the results to be contrasted with those in FIG. \ref{fig2}. A maximum, but still finite, delay is obtained for the earliest switch at $\Gamma t_{sw}=0$, its value $\Gamma t_{end} = \ln (3+\sqrt{5})/2 \approx 0.9624$ being a little less than double that of $\Gamma t_0$. A simple, analytical expression describes the curve in FIG. \ref{fig3}. With $x = \exp (-\Gamma t_{sw}), y = \exp (-\Gamma t_{end})$, we have 

\begin{equation}
y(x) = \frac{3-\sqrt{9-24 x+20 x^2}}{2(2-x)}.
\label{eqn5}
\end{equation}
The value of $\Gamma t_0$ in Eq.~(\ref{eqn4}) corresponds to the root $y=x=2-\sqrt{2}$.

In summary, we have shown that a simple local unitary operation that can be carried out on both qubits of an entangled pair changes the subsequent evolution of their entanglement. For mixed states under conditions which lead to a loss of that entanglement at finite time, termed sudden death, such an operation can either hasten or delay that death, depending on the time at which it is carried out. There is a critical time before which the operation can even completely avert the sudden death of entanglement. When the local transformation is done at only one of the qubits, sudden death is hastened or delayed but not averted completely.   

\acknowledgements

One of us (ARPR) thanks the Theoretische Quantenphysik group at the Technische Universit\"{a}t Darmstadt, for its hospitality during the course of this work. M. Ali acknowledges financial support by the Higher Education Commission, Pakistan, and the Deutscher Akademischer Austausch Dienst.

\end{document}